\documentstyle[12pt]{article}
\begin{document}
\pagestyle{plain}
\begin{titlepage}
\vspace*{3cm}
\begin{center}
{\Large \bf The Magnetization of the 3D Ising Model
}\\[2ex]
{\bf
A. L. Talapov\dag\ and  H. W. J. Bl\"ote\S}
\\[1ex]
{\sl \dag\ Landau Institute for Theoretical Physics,\\ GSP-1 117940
Moscow V-334, Russia\\
\S\ Faculty of Applied Physics, Delft University of Technology,\\
Lorentzweg 1, 2628 CJ Delft, The Netherlands
}
\end{center}

\noindent
PACS. 05.50 -- Lattice theory and statistics; Ising problems\\

\begin{center}
{\bf Abstract}
\end{center}

We present highly accurate Monte Carlo results for simple cubic Ising
lattices containing up to $256^3$ spins. These results were obtained
by means of the Cluster Processor, a newly built special-purpose
computer for the Wolff cluster simulation of the 3D Ising model.
We find that the magnetization $M(t)$ is perfectly described by
$M(t)=(a_0-a_1 t^{\theta} - a_2 t) t^{\beta} $, where
$t=(T_{\rm c}-T)/T_{\rm c}$, in a wide temperature range
$0.0005 < t < 0.26 $.  If there exist corrections to scaling with
higher powers of $t$, they are very small.
The magnetization exponent is determined as $\beta=0.3269$ (6).
An analysis of the magnetization distribution near criticality yields
a new determination of the critical point:
$K_{\rm c}=J/k_B T_{\rm c}=0.2216544$, with a standard deviation of
$3\cdot 10^{-7}$.
\end{titlepage}

We consider the 3D Ising model on the simple cubic lattice, with
nearest-neighbor interactions $J$, at a temperature $T$ and a coupling
strength $K=J/k_BT$.  At criticality, the spontaneous magnetization
vanishes with a singularity $M(t) \propto t^{\beta}$, where
$t=(T_{\rm c}-T)/T_{\rm c}=(K-K_{\rm c})/K$ parametrizes the
distance to the critical point. However, this law applies only in the
limits of infinite system size and  $t \rightarrow 0$.
Even for the infinite system there are corrections due to
nonzero values of $t$:
\begin{equation}
M(t)=f_t(t) t^{\beta}
\label{Mt}
\end{equation}
where $f_t$ is some function of $t$, finite at $t=0$.

For the 2D Ising model this function is known exactly, and it is analytic.
However, in the 3D case, $f_t$ is not analytic at $t=0$. The leading
terms of its expansion near $t=0$ are
\begin{equation}
f_t(t) \approx a_0 - a_1 t^{\theta}
\label{ft0}
\end{equation}
where $\theta \approx 0.5$ is Wegner's correction-to-scaling
exponent~\cite{Wegner}.

Generally one would expect that there exist many more terms, containing
higher powers of $t$, in the expansion of $f_t$. Quite remarkably, we
find that it is sufficient to add only one term, $(-a_2 t)$, to
Eq.~(\ref{ft0}), in order to describe $M(t)$ of the simple cubic 3D
Ising model with very high accuracy. This does not only apply
close to $t=0$; it holds in a wide range of $t$.

We found this intriguing fact using the Cluster Processor
(CP)~\cite{3DSPC}.  The CP implements the cluster Wolff
algorithm~\cite{Wolff} in hardware, for 3D simple cubic Ising models
with nearest-neighbor interactions and periodic boundaries. Its memory
and speed are sufficient to simulate Ising systems containing up to
$256^3$ spins. The CP was checked to give correct results for 2D Ising
systems. Moreover the CP data are also consistent with earlier very
accurate simulations on $16^3$ and $32^3$ lattices \cite{Univers}.
In the CP, the system size can take the 5 values $16$, $32$, $64$, $128$,
$256$ along each spatial direction. But the present work is
restricted to systems with the same size $L$ in all three directions.

To determine $f_t$, one should eliminate the influence of the finite
system size. This can be easily achieved by a comparison of data for
different sizes. The corrections due to finite $L$ can be described
by a scaling function $f_L$:
\begin{equation}
M(t)=f_L(Lt^{\nu}) t^{\beta}
\label{ML}
\end{equation}
where the exponent $\nu$ describes the divergence of the correlation
length when the critical temperature is approached.
The relation Eq.~(\ref{ML}) is valid in the limit $t \rightarrow 0$.
The argument of the function $f_L$ is proportional
to the ratio of the finite size and the correlation length. Thus,
$f_L$ is expected to be a constant for large values of its argument.

The meaning of $M(t)$ in Eq.~(\ref{ML}) has to be made more precise, since
the average magnetization of a finite system vanishes. Instead, we may
define a nonzero expectation value in terms of the absolute value of $M$:
\begin{equation}
M_m=\langle |M| \rangle
\label{Mm}
\end{equation}
We can also find $M(t)$ from the spin-spin correlation function, using
the relation
$M^2=\langle S(0)S(\infty) \rangle$.
For a finite lattice this infinite distance
can be replaced by $L/2$, half way to the periodic images of the spins.
Therefore we use the following expression as the second definition of
the magnetization:
\begin{equation}
M_c=\langle S(0)S(L/2) \rangle^{1/2}
\label{Mc}
\end{equation}
Finally, the magnetization can be defined as
$M_2=\langle M^2 \rangle^{1/2}$.
This quantity has been studied in \cite{Miyashita}
for the 2D Ising model and in \cite{Ito} for the 3D case.
In agreement with \cite{Miyashita,Ito} we found that
$M_2$ is strongly affected by finite-lattice
effects even for large $t$, so we do not use $M_2$ in this
paper.

The scaling functions $f_L$ are different for
$M_m$ and $M_c$. This helps to determine the ranges of
$t$ where the finite-size effects become important.

Our data show, that for a given $L$, the finite-size corrections to
$M_m$ and $M_c$ are smaller than the
error bars for $t > t_L$. For $L=32$ this is illustrated by Fig.~1,
which clearly demonstrates that $M_m$ and $M_c$ coincide for $t>t_{32}$,
and $t_{32}$ is about $1.5\times 10^{-2}$. From Fig.~1
one can see that for $t$ just below $t_L$ the corrections to
$M_m$ and $M_c$ have different signs, which facilitates determination
of $t_L$.

The value of $t_L$ can be estimated in the following way.
The finite-lattice effects become important when the correlation
length $\xi \propto t^{-\nu}$ is comparable with $L$. Therefore
\begin{equation}
a t_L^{-\nu} = L
\label{tL}
\end{equation}
where $a$ is a numerical coefficient. Taking into account $\nu=0.63$
\cite{Univers}, we get  $a \approx 2$, which seems reasonable.
According to Eq.~(\ref{tL}), a doubling of $L$ decreases $t_L$ by a factor
$2^{1/\nu} \approx 3$. This is in agreement with Fig.~2, which shows
the normalized $M_m(t)$ results for three different lattice sizes.

To study $f_t(t)$ we used data for lattice sizes ranging
from $32^3$ to $256^3$. Only data for $t>t_L$, where $M_m$ and $M_c$
coincide, were taken into account. The results for $M(t)$ are shown in
Fig.~3, using a logarithmic scale on both axes.
Close to the critical point, for $t < 0.02$, the plot seems to be linear.
However, attempts to approximate the data by
$t^{\beta}P(t)$,
where $P(t)$ is an arbitrary polynomial in $t$, were
not successful. Fig.~4 shows the poor result of such an attempt.
It describes the ratio of the simulation data to
\begin{equation}
M_{\rm int}(t)=t^{\beta}(p_0+p_1 t +p_2 t^2 + p_3 t^3)
\label{mint}
\end{equation}
The exponent $\beta$ and the coefficients $p_i$ were determined
by the least-squares method, in order to describe the simulation data
as closely as possible by Eq.~(\ref{mint}). Nevertheless, the differences
between the CP data and the approximation by Eq.~(\ref{mint})
are much larger than the error bars.

This suggests that we should use the form Eq.~(\ref{ft0}) instead to
describe $f_t(t)$.
Thus we wrote $f_t(t)$ as a polynomial in $t^{1/2}$
\begin{equation}
f_{t1}=a_0 - a_1 t^{1/2} - a_2 t
\label{sqr}
\end{equation}
Even the first attempt to approximate the simulation data as
$t^{\beta}f_{t1}$
with $\beta=0.3267$, taken from \cite{Univers}, and $a_0,a_1,a_2$
regarded as free parameters, was very successful.
But the high statistical accuracy of the CP data allows the use of
even more adjustable parameters, and we supposed that
the magnetization can be described by
\begin{equation}
M_0(t)=(a_0 - a_1 t^{\theta} - a_2 t ) t^{\beta}
\label{m0}
\end{equation}
in the interval $0.26 >t >0.0005$.
The five parameters $\beta$, $\theta$ and $a_i$ were determined by the
least-squares method. To estimate the influence of the uncertainty in
$K_{\rm c}$, we fitted the parameters not only for our best estimate
$K_{\rm c}=0.2216544$ (see below), but also for the lower and upper
limits of $K_{\rm c}$, defined by one standard deviation
($3\times10^{-7}$) of the critical coupling. The results are shown in
Table 1.
\begin{table}[htbp]
\caption{Results of least-square fits of the magnetization data obtained
  by the CP. These fits were made for three choices of the critical
  coupling $K_{\rm c}$.}
\label{tab:fits}
\begin{center}
\begin{tabular}{c|c|c|c|c|c|c}
\hline
\hline
$K_{\rm c}$&$\beta$ & $\theta$ & $a_0$   & $a_1$   & $a_2$   &$\chi^2$\\
\hline
0.2216544 &0.3269(3)&0.508(15) &1.692(4) & 0.344(6)&0.426(11)&0.844\\
0.2216541 &0.3274(3)&0.490(15) &1.698(4) & 0.340(5)&0.436(10)&0.859\\
0.2216547 &0.3265(3)&0.528(15) &1.686(4) & 0.348(7)&0.414(12)&0.848\\
\hline
\hline
\end{tabular}
\end{center}
\end{table}
\noindent
The last column characterizes the quality of the least-squares
approximation. It is defined as
$
\chi^2=\frac{1}{N-n} \sum_{i=1}^{N} \left(
\frac{M(t_i)-M_0(t_i)}{\sigma(t_i)} \right)^2
$,
where $n=5$ is the number of fitted parameters, $N=45$ is the number of
magnetization data points, and the $\sigma(t_i)$ are the standard deviations
of the magnetization data $M(t_i)$.
The minimum of  $\chi^2$ as a function of $K$ appears to occur near
$K_{\rm c}=0.2216544$ as determined from an analysis of the magnetization
distribution near the critical point (see below). This agreement
between two different approaches suggests that our scaling formulas
are adequate.

>From the data in the second column we estimate the magnetization exponent
as $\beta=0.3269$ (6), which is in a good agreement with earlier values,
see e.g. Ref.~\cite{Univers} and references therein.

The third column indicates that  $\theta=0.508$ (25), supporting
earlier results obtained by means of an $\epsilon$-expansion
analysis~\cite{LGZJ2}, series expansions~\cite{NickR} and a finite-size
scaling analysis of three different Ising models~\cite{Univers}.

It should be emphasized that the accuracy of Eq.~(\ref{m0})
is much better than one might expect from the quoted standard
deviations of the formula parameters.
To utilize this property, we present here a very accurate empirical
approximation for the spontaneous magnetization of the simple cubic
Ising model in the region $0.26 > t >0.0005$:
\begin{equation}
M_0(t)=t^{0.32694109}(1.6919045 - 0.34357731 t^{0.50842026}-0.42572366 t)
\label{mapprox}
\end{equation}
where $t=1-0.2216544 \, k_BT/J$.

In the range of $t$ between $0.24$ and $0.17$ the results of
Eq.~(\ref{mapprox}) numerically coincide with the Pad\'e approximant of
Ref.~\cite{EF} within $10^{-5}$, while
for smaller $t$ the approximation Eq.~(\ref{mapprox}) is superior.

The ratio of the CP data to $M_0(t)$ is shown in Fig.~5. The simulation data
coincide with $M_0(t)$ within the error bars. No systematic deviation
of the CP data from $M_0(t)$ can be found.

The function $f_t$ may also contain a term proportional to
$t^{2 \theta}$. Because $\theta$ is very close to $0.5$,
the simulation accuracy is not sufficient to distinguish the term
$(a_2 t)$ in Eq.~(\ref{m0})
from the sum $(a_{21} t + a_{22} t^{2 \theta})$.

The CP was also used for some additional calculations
close to the critical point. We obtained the dimensionless
ratio $Q=\langle m^2\rangle^2/\langle m^4\rangle$, related to
the Binder cumulant~\cite{Binder}. The $Q$ values are shown in
Table~\ref{CPqdata}.
\begin{table}[htbp]
\caption{Numerical results for the dimensionless ratio $Q_L= \langle
  m^2\rangle_L^2 / \langle m^4\rangle_L$ for finite three-dimensional
  Ising models close to the critical point.
  Also shown is the total number of
  Wolff clusters flipped by the CP to obtain each numerical result.}
\label{CPqdata}
\begin{center}
\begin{tabular}{||r|c|l|r||}
\hline
 $L$  &   $K$     &   $ Q_L $      &    $\#$ clusters         \\
\hline
   16 & 0.2216530 &  0.63381   (5) & $1   \times10^9$      \\
   32 & 0.2216530 &  0.62883  (10) & $3   \times10^8$      \\
   64 & 0.2216530 &  0.62469  (79) & $2.5 \times10^7$      \\
   64 & 0.2216545 &  0.62670  (25) & $2.5 \times10^8$      \\
  128 & 0.2216530 &  0.62189 (168) & $1   \times10^7$      \\
  128 & 0.2216545 &  0.62603  (84) & $4   \times10^7$      \\
  256 & 0.2216530 &  0.61555 (180) & $2.6 \times10^7$      \\
\hline
\end{tabular}
\end{center}
\end{table}

These data were combined with those available from Ref.~\cite{Univers} in
order to determine the critical point more accurately. Near $T_{\rm c}$,
the bulk correlation length satisfies $\xi \gg L$, so that $Lt^{\nu}$,
and hence $(K-K_{\rm c}) L^{y_t}$, is small. Thus $Q_L(K)$  can be
expanded as
\begin{equation}
  Q_L(K)= Q + q_1 (K-K_{\rm c}) L^{y_t}+
  q_2 (K-K_{\rm c})^2 L^{2y_t}+
  b_1 L^{y_{\rm i}} + b_2 L^{y_2}
\label{expan}
\end{equation}
The renormalization exponents $y_t$  and $y_{\rm i}$
are related to $\nu$ and $\theta$ as $y_t$=$\nu^{-1}$ and
$y_{\rm i}$=$- \theta/ \nu$.
The substitutions of the values $y_t=1.587(2)$, $y_{\rm i}=-0.82(6)$
and $y_2=-1.963(3)$, taken from \cite{Univers}, into Eq.~(\ref{expan}) was
found to describe the combined data satisfactorily. A least-squares fit
was used to determine $K_{\rm c}$.
For system sizes $L\geq5$ the same fit as in \cite{Univers}
yielded $K_{\rm c}=0.2216544(3)$, where the standard error includes
the uncertainty in the input parameters. As a final estimate we quote
$K_{\rm c}=0.2216544(6)$ with an error of two standard deviations, in order
to account for a possible bias introduced by our choice of the form of
Eq.~(\ref{expan}). The additional data in Table 2 permitted a clear
improvement of the accuracy in comparison with Ref.~\cite{Univers}.

We calculated the irrelevant exponent $y_{\rm i}$ in two ways.
First, it can be obtained as a product of $\theta=0.508(25)$,
found from Table (1), and
$y_t$=$1.587(2)$ \cite{Univers}. The result is $y_{\rm i}=-0.81$ (4).
Another possibility is to include $y_{\rm i}$ as a free parameter in the
$Q_L(K)$ fitting procedure. This yields $y_{\rm i}=-0.83$ (9), in agreement
with the result given above, and with earlier results~\cite{Univers},
based on an analysis including three different Ising models. Our
estimates of $y_{\rm i}$ agree with $\epsilon$-expansion
results of \cite{LGZJ2}.

Furthermore, we have combined the new data for $\langle m^2 \rangle$
near the critical point with those obtained in Ref.~\cite{Univers} for
the nearest-neighbor model with system sizes up to $L=40$. The analysis
was based on the expected scaling behaviour of the susceptibility
(see Ref.~\cite{Univers})
$$
\begin{array}{l}
  L^d \langle m^2 \rangle =
  c_0 + c_1 (K-K_{\rm c}) + \cdots\\
  \qquad {} + L^{2y_h-d} \left[
  d_0 + d_1(K-K_{\rm c})L^{y_t} + d_2(K-K_{\rm
    c})L^{2y_t} + g_1 L^{y_{\rm i}} + g_2 L^{2 y'} \right]
\end{array}
$$
with $K_{\rm c}=0.2216544$ and $y'=-2.1$.
The renormalization exponent $y_h$ is related to the magnetic
susceptibility critical exponent $\gamma$: $2y_h-d=\gamma/ \nu$.
The analysis for $L\geq5$ yielded
$y_h =  2.4808$ (16) where we again quote
a two-sigma error. This result is in a good agreement with
Ref.~\cite{Univers} and references therein, and with Ref.~\cite{BHHMS}.
\\[1.5ex]
{\bf Acknowledgments}:
We are much indebted to  A.F. Bakker, A. Compagner, J.R. Heringa,
A. Hoogland, E. Luijten, W. Selke, and L.N. Shchur
for their cooperation and for sharing their valuable insights.

This work was partially supported by grants 07-13-210 of NWO,
the Dutch Organization of Scientific Research, INTAS-93-0211,
M0Q300 of ISF, the International Science Foundation and
93-02-2018 of RFFR, the Russian Foundation for Fundamental Research.
\\[1.5ex]

\newpage

\noindent
{\bf Figure captions}\\[1ex]

\noindent
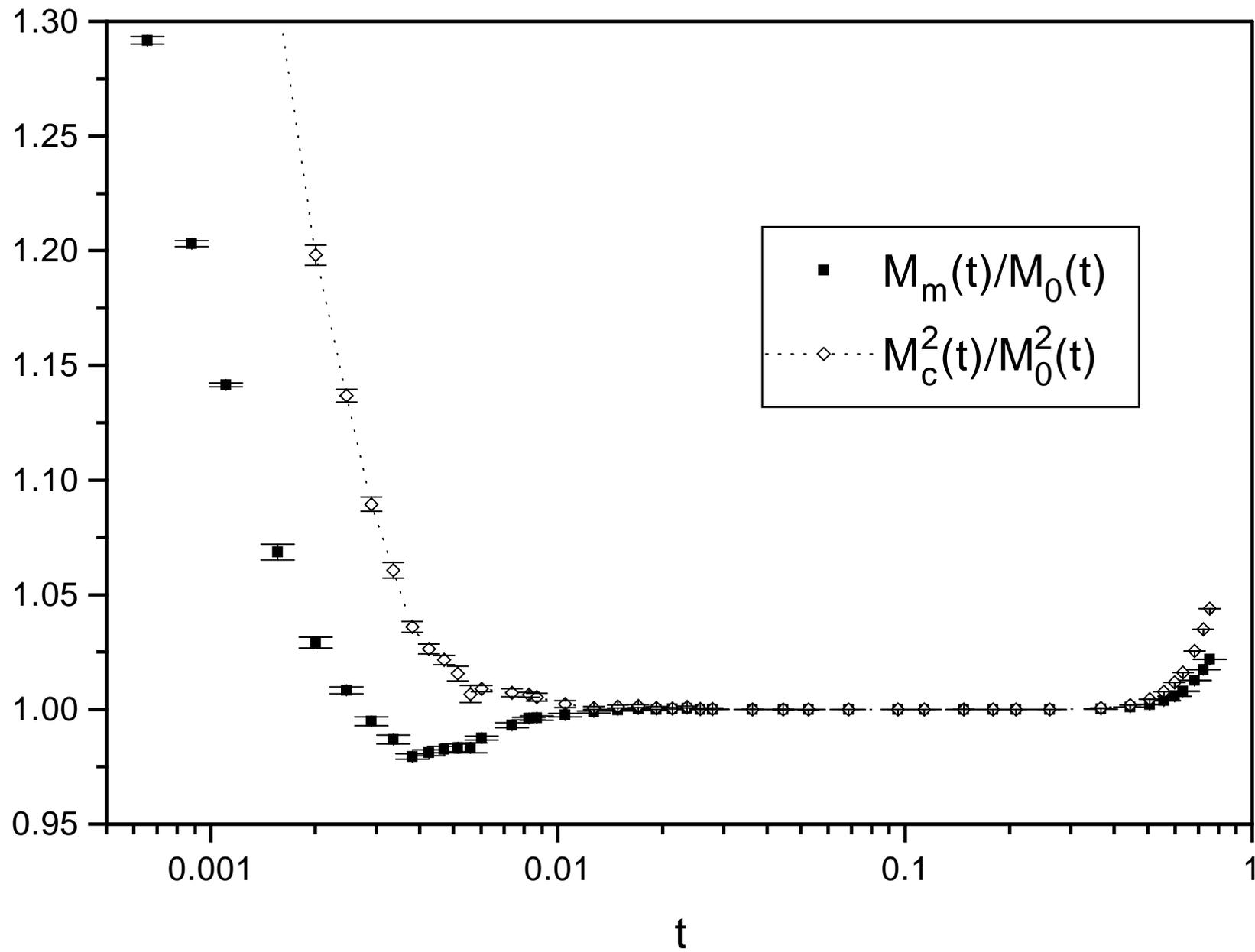
{\bf Fig.~1:} Normalized magnetization $M_m(t)$ and the correlation
function $\langle S(0)S(L/2) \rangle \equiv M_c^2$ for the $32^3$ lattice.
$M_0(t)$ is given by Eq.(\ref{m0}).
The normalization makes it possible to expand the scale so that the small
deviations of $M(t)$ from $M_0(t)$ become visible.
$\langle S(0)S(L/2) \rangle$ is close to $M_0^2(t)$,
and $M(t)$ is close to $M_0(t)$ for
$t>t_{32} \approx 0.015$.
For $t > 0.26$ the deviations of the CP data from $M_0(t)$ grow rapidly.
In this low-temperature range the simulation data are in an excellent
agreement with the Pad\'e approximant of Ref.~\cite{EF}.
\\[1ex]

\noindent
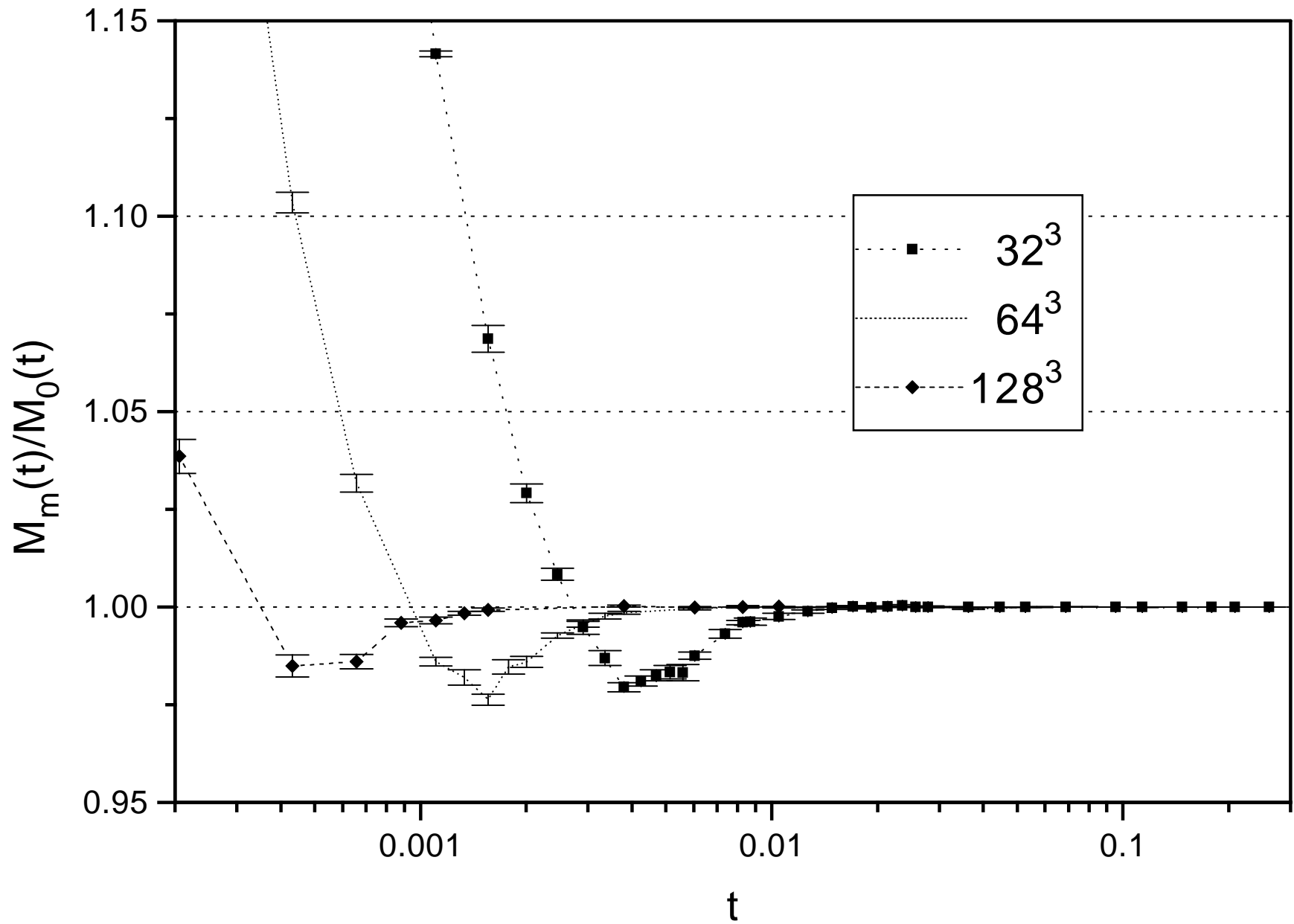
{\bf Fig.~2:} Normalized magnetization $M_m(t)$ for 3 different lattice
sizes $L=32$, 64, and 128. The small-$t$ behavior displays the
characteristics of the finite-size scaling function $f_L$ associated
with $M_m$.
\\[1ex]

\noindent
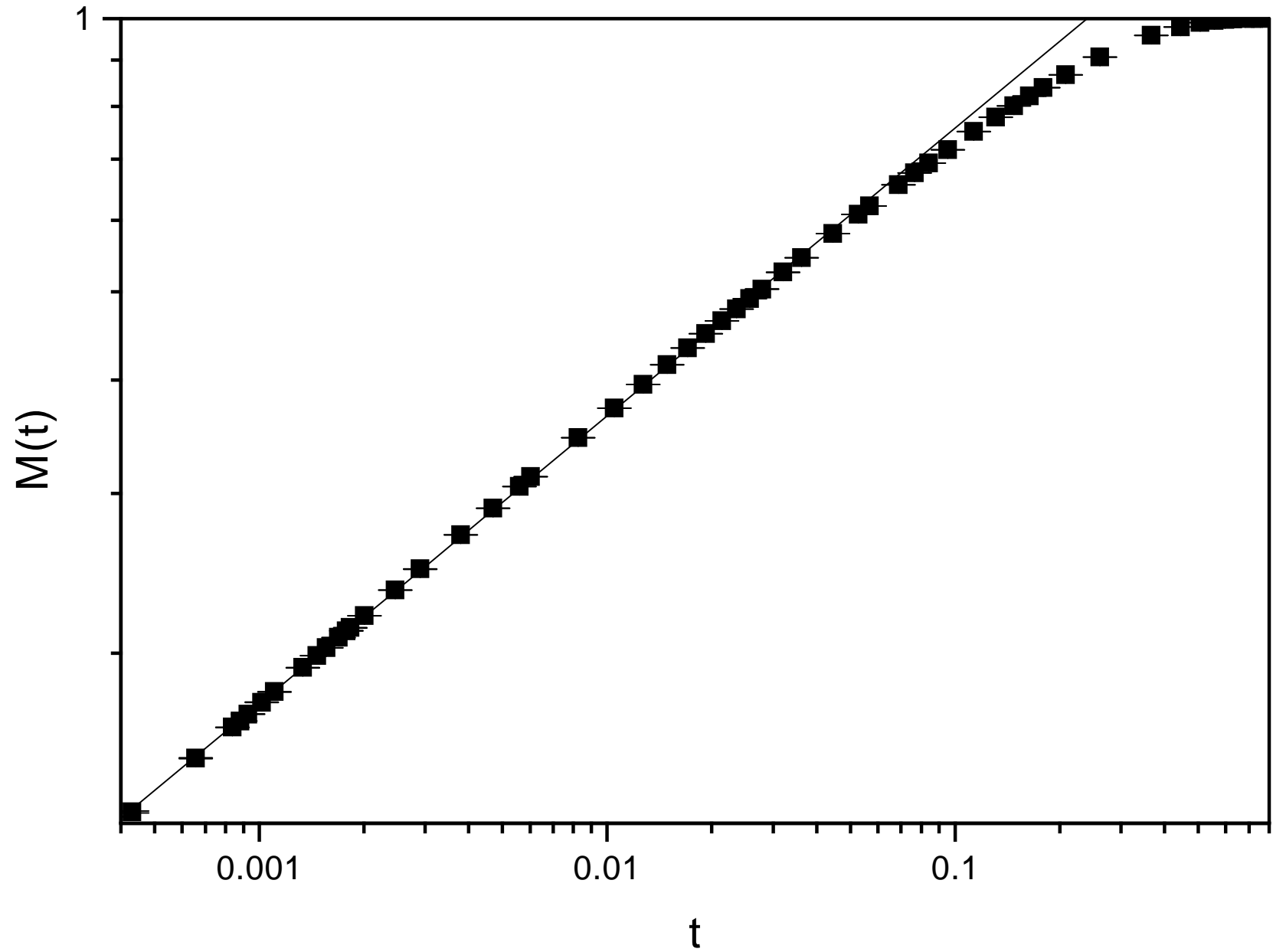
{\bf Fig.~3:} Spontaneous magnetization of the 3D Ising model as a
function of $t$, with a logarithmic scale on both axes.
The statistical errors in these data points are indicated, but they appear
only as single horizontal bars, because the errors are below the resolution
of this figure.
The data shown here apply to different lattice sizes, and were taken in
those ranges of $t$ were finite-lattice effects are negligible:
they describe the infinite system magnetization.
\\[1ex]

\noindent
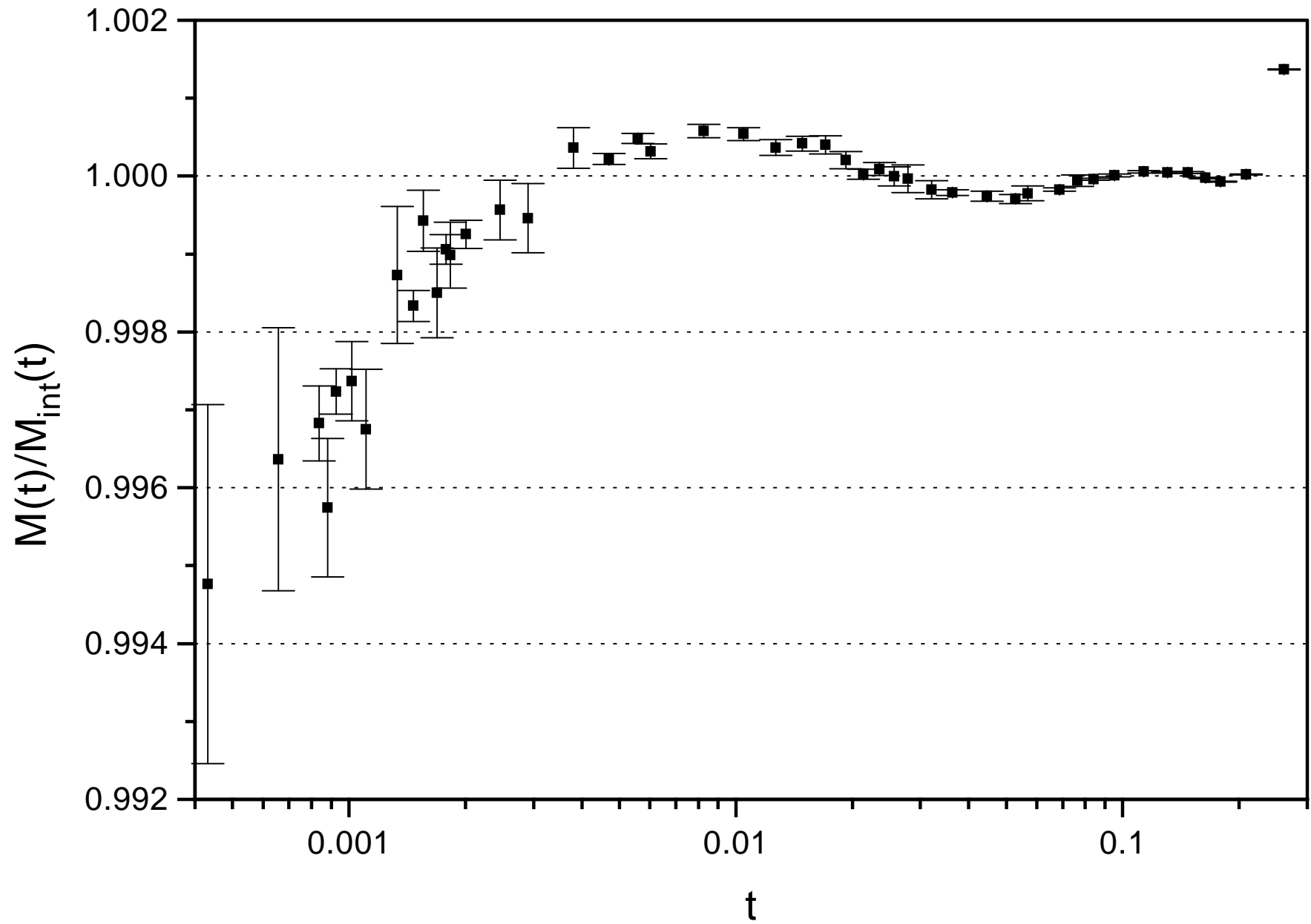
{\bf Fig.~4:} Ratio of magnetization data to the approximation
$M_{\rm int}(t)$ given in Eq.~(\ref{mint}), where the function $f_t(t)$ was
supposed to contain only integer powers of $t$.
This Figure demonstrates that, without the Wegner correction to scaling,
even a 5-parameter fit according to Eq.~(\ref{mint}) does not describe
the CP data properly. In this equation we used the same number of
adjustable parameters as in $M_0(t)$ (see Eq.~\ref{m0}), but, as can be
seen by comparing Figs.~4 and 5, the latter approximation is incomparably
better.
\\[1ex]

\noindent
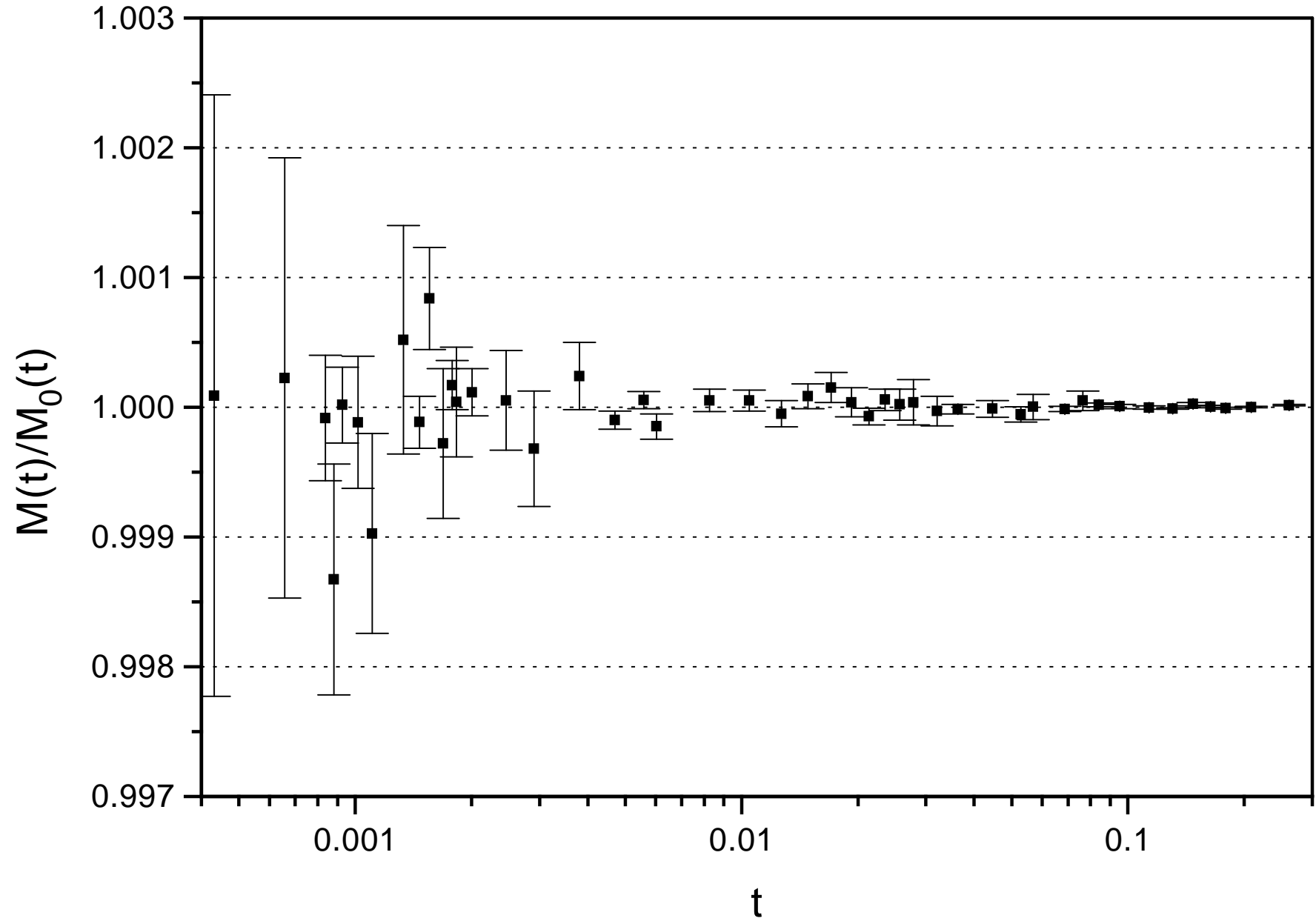
{\bf Fig.~5:} Normalized magnetization data, describing the infinite 3D
Ising system. This Figure demonstrates that the expression Eq.~(\ref{m0})
for $M(t)$ agrees with the simulation data within the statistical errors.
The relative accuracy of the formula is as high as $10^{-3}$
for $t \approx 10^{-3}$, and better than $10^{-4}$ for $t > 0.01$.
This picture combines all the magnetization data, obtained with different
precisions and for different lattice sizes. Long simulation times were
required for some of these points in order to obtain such small error bars.
\\[1ex]

\end{document}